# Laparoscopic real-time *en-face* differential optical topography of near-surface heterogeneity by using an applicator probe with the maximum-density radially alternating illumination and detection fiber-channels

Daqing Piao,[1*] Kenneth E. Bartels,[2] G. Reed holyoak,[2] Sanjay Patel [3]

[1]*School of Electrical and Computer Engineering, Oklahoma State University, Stillwater, OK, 74078*
[2]*Department of Veterinary Clinical Sciences, Center for Veterinary Health Sciences, Oklahoma State University, Stillwater, OK, 74078*
[3]*Department of Urology, University of Oklahoma Health Sciences Center, Oklahoma City, OK, 73104*

*Corresponding author: daqing.piao@okstate.edu

**We demonstrate a laparoscopic applicator probe and a method thereof for real-time *en-face* topographic mapping of near-surface heterogeneity for potential use in intraoperative margin assessment during minimally invasive oncological procedures. The probe fits in a 12mm port and houses at its maximum 128 copper-coated 750μm fibers that form radially alternating illumination (70 fibers) and detection (58 fibers) channels. By simultaneously illuminating the 70 source channels of the probe that is in contact with a scattering medium and concurrently measuring the light diffusely propagated to the 58 detector channels, the presence of near-surface optical heterogeneities can be resolved in an *en-face* 9.5mm field-of-view in real-time. Visualization of a subsurface margin of strong attenuation contrast at a depth up to 3mm is demonstrated at one wavelength at a frame rate of 1.25Hz. © 2016.**

Laparoscopic and robot-assisted oncological procedures increasingly become the preferred approaches for patients and surgeons, specifically in urologic surgery. When compared to open surgery, oncological procedures done laparoscopically are impaired by the difficulty of obtaining timely intraoperative pathology consultation [1], stereotactically confined field-of-view and decreased (pure laparoscopic surgery) or nonexistent (robot-assisted laparoscopic surgery) tactile sensation [2]. With these challenges, surgeons operating laparoscopically often have to rely upon subjective visual cues to guide excision of tumors (i.e. adequate resection of renal tumor during partial nephrectomy (PN)) and avoide tumor violation or iatrogenic injury to critical adjacent tissues (i.e. identifying the prostatic capsule and preserving peri-prostatic nerve tissues during radical prostatectomy (RP)).

Among laparoscopic imaging tools available for intraoperative tissue assessment, drop-in ultrasound [3] is particularly useful for localizing tumor depth; however, the tumor depth evaluated in the transverse ultrasound view does not directly indicate the tumor margin over the lateral field-of-view (FOV) whereupon the resection is performed. Laparoscopic wide-field imaging of Firefly fluorescence has shown to enhance the outcome of PN [4]. Wide-field imaging of the surface fluorescence is done in a lateral FOV that is ideal for direct navigation guidance; yet, as the image mainly reflects superficial fluorescence emission [5], tissue heterogeneity beneath the surface due to invasive lesions could be underestimated or missed [6].

A laparoscopic imaging tool to allow assessment of the tumor margin in PN or identification of the prostatic capsule amid peripheral nerve tissues in RP is desirable to probe tissue contrast (due to absorption, scattering, and fluorescence, etc.) at a few millimeters depth [7] as the resection proceeds, visualize over the lateral view for resection guidance, have a non-microscopic FOV adequate for rapid yet detailed survey of the resection site, and form the image in real-time. Probing light diffusely propagated through tissue provides sub-surface sensitivity as with diffuse or laminar reflectance imaging [8, 9] or spatial frequency domain imaging [10], but projecting these modalities laparoscopically to sample subsurface tissue heterogeneity has been challenging, and additionally the image formation with some of these modalities could involve intense computation that may be costly to the intraoperative time-frame.

In this paper we demonstrate a new laparoscopic contact-based optical imaging technology, *en-face* differential optical topography (*en-face* DOT), which performs real-time visualization of subsurface tissue heterogeneity within a depth up to 3mm and over a 9.5mm diameter FOV with a modest mm-level lateral resolution. This *en-face* DOT device combines well-defined depth-sampling of separate source and detector channels and spatial resolution rendered by the maximum-density fiber channels. *En-face* DOT uses a two-dimensional fiber-array as is common to diffuse reflectance imaging or tomography [8, 9]. But *en-face* DOT operates differently in two aspects: (1) the fiber-array is set at the maximum density and (2) the signals originating from all source channels are acquired compositely, i.e., no multiplexing of the source channels. This all-channels-ON mode may seem like using an imaging-fiber [11] (typically has a 1mm or less diameter so the FOV when in contact with tissue is 1mm or less), but there is a fundamental difference. An imaging-fiber has fully coherently ordered fiber channels, and the individual μms-size fiber is used for both illumination and light-detection thereby the imaging essentially samples the surface. In comparison, the *en-face* DOT probe has semi-coherently ordered fiber channels-----the individual 100s-μm size fiber acts as either the source-channel or detector-channel, but not both, to render depth-sampling that is unavailable to an image-fiber. This unique maximum-density semi-coherent fiber-array combined with all-sources-ON operation directly (thus real-time) maps the lateral distributions of optical sources of contrast, at a 1~2 mm sampling depth, over an *en-face* FOV sized by the applicator probe, with a lateral resolution set by the individual fiber. Although the present demonstration of real-time visualization is limited to the qualitative contrast of continuous-wave light attenuation at a single-wavelength, the method can be extended to quantitatively probing the spectral or fluorescence heterogeneity, and potentially time-resolved detections.

The principle of *en-face* DOT is illustrated in Fig. 1. Figure 1(A) depicts 3 sets of side-by-side placed fibers that form three source-detector (SD) pairs with the shortest possible SD distance. Each of such side-by-side SD pair is sensitive to only a shallow region directly underneath the pair, e.g., the SD pair 1 is minimally sensitive to a deep-seated anomaly 1, the SD pair 2 is moderately sensitive to a nearby anomaly 2, and the SD pair 3 is maximally sensitive to a beneath-the-pair anomaly 3. Each side-by-side SD pair thus has a localized lateral and depth sensitivity defined by the fiber size. When these side-by-side SD pairs are stacked along one dimension to form a maximum-density array with alternating source and detector channels under uniform illumination as is shown in (B), and when all source channels are illuminated simultaneously, the composite signal reaching each detector and originating from all sources is dominated by those originating from the sources the closest to the detector. Thus the sensitivity of each side-by-side SD pair to the shallow volume below it is nearly maintained, which renders the one-dimensional composite-signal profile to directly map the lateral distribution of the shallow tissue heterogeneity. For the one-dimensional maximum-density array in (B) that has a total of M source and N detector positions (N=M+1 or M=N+1), the tissue length in contact with the array as marked by the horizontal solid arrow can be divided into (M+N) segments, each representing a column beneath a fiber. The complete set of signals acquired at the N detector positions and would have been acquired at the M source positions (if the sources and detectors are swapped and the signal is scaled by M/N) thus directly profiles the shallow tissue heterogeneity with a lateral resolution set by the fiber size. When a two-dimensional array with the maximum fiber density is used as in (C), tissue anomaly at a depth sensitive to neighboring side-by-side SD pairs is directly mapped over the *en-face* plane. Spectral or fluorescence measurements can be readily achieved by using such a maximum-density fiber array to probe beneath-surface endogenous or exogenous contrast. Note that the source/detector channels in (C) alternates only along one direction, in contrary to a fully mixed source/detector channels in a 2cm×2cm square applicator for rapid data acquisition but non-real-time image formation [12]. What this Letter demonstrates is real-time image formation, using a circular probe, with radially alternating source and detector channels.

We have prototyped a laparoscopic applicator probe for demonstrating *en-face* DOT in an FOV of 9.5mm in diameter, using a standard blunt-tip trocar fitting a 12mm stability sleeve port, as shown in Fig. 2A and 2B. The optical tip of the trocar was carefully removed to open up the stainless steel stem (Fig. 2C) for housing optical fibers. A total of 128 copper-coated 750μm fibers (Oxford Electronics, IR600/660, core/cladding/coating 600/660/750nm, 0.22NA) were enclosed by the tip-removed stainless steel stem (Fig. 2D). These fibers form approximately 6 circles concentric to the approximate center of the probe occupied by a fiber. A total of 70 fibers forming the odd number of circles (the center fiber is counted as circle 1) are used as the source channels. The remaining 58 fibers forming the even number of circles (the 6 fibers in direct contact with the center fiber are counted as circle 2) are used as the detector channels. Due to limitations in fabrication process, the fibers at the periphery of the probe are less evenly distributed comparing to those in the inner part of the probe, as shown in Fig. 2E and 2F where the 70 source and 58 detector channels are separately illuminated for visual inspection.

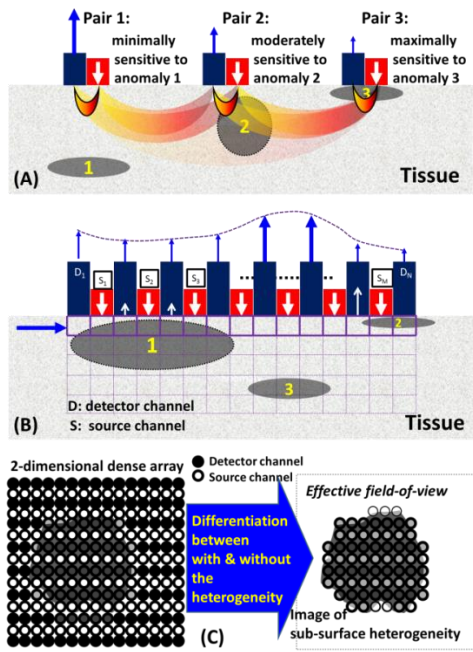

**Fig. 1.** (A): For a linear array as shown, each detector is most sensitive to the anomaly present in the light path between the detector and its closest source. This 1D fiber array thus resolves a shallow anomaly with a lateral resolution of approximately twice of the fiber size. (B) The 2D array of side-by-side fibers with alternating rows of source and detector channels resolves a shallow anomaly in an en-face view, at lateral resolutions of approximately one and two times of the fiber size in the shown horizontal and vertical directions, respectively.

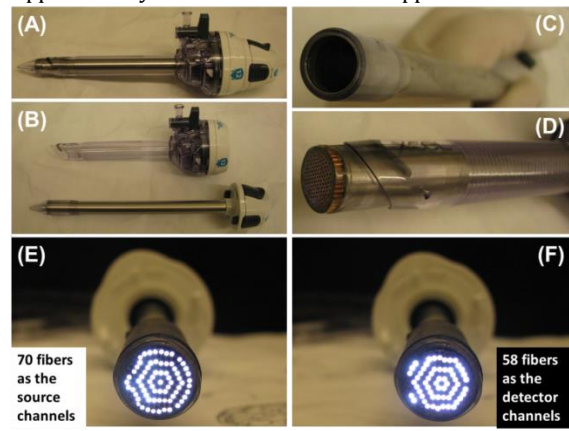

**Fig. 2.** (A) A 12mm bladeless trocar in the stability sleeve. (B). The bladeless trocar separated from the stability sleeve. (C) The stainless steel trocar casing after removing the bladeless tip. (D) The trocar-based probe housing 128 copper-coated fibers of 750um in diameter is placed in the stability sleeve. (E) 70 source fibers occupying the center and the even number of the rings of the fiber positions. 6). 58 detector fibers occupying the odd number of the rings of the fiber positions.

The system configuration for *en-face* DOT using the laparoscopic prototype probe shown in Fig. 2 is illustrated in Fig. 3. The 70 fibers as the source channels and 58 fibers as the detector channels are separately bundled for coupling to the instrument console. Due to shortage of the copper-coated fiber for the prototype development, the two bundled tails of the probe were short and thus rigid. The total length of the fiber probe from the distal tissue-contacting facet to the rigid proximal instrument end is approximately 2 ft. The bundle of the 70 source fibers is evenly illuminated by a laser diode or a light emitting diode (both sources were tested whereas the data presented were acquired by using a 785nm laser diode, a 525B Laser Diode Driver and a TED 200C Temperature Controller, Thorlabs Inc, Newton, NJ) after collimating by a lens (C220TME-B, Thorlabs Inc,. Newton, NJ) and passing through a diffuser (10deg 25mm, Edmund Optics, Barrington, NJ). The bundle of the 58 detector fibers is imaged onto a camera (PointGrey GS3-U3-28S4M-C, 1928×1448 pixels) using a 4X microscope objective lens and a relay lens (100mm focal length). Because the instrument ends of the laparoscopic probe are rigid, the optical modules for coupling fiber-delivered light into the probe and the probe-detected light onto the camera have to be fixed at a frame-structure that also holds the probe vertically. An XYZ translation stage is used to position an object underneath the probe for testing. When the probe is in contact with a 1% intralipid solution corresponding to a reduced scattering coefficient of 1mm$^{-1}$ and an absorption coefficient of 0.0025mm$^{-1}$ at 785nm, one frame of raw image data is acquired in 8ms at a camera gain of 0dB, using the vendor-provided FlyCapture 2.8 interface. The raw image data when reshaped off-line on a CORE i5 processor running on Windows 7 for image formation takes approximately 78ms. A streamlined interface developed in LabView incorporating MATLAB scripts of image reshaping algorithm currently runs at 1.25Hz. Video-rate imaging is possible with future optimization.

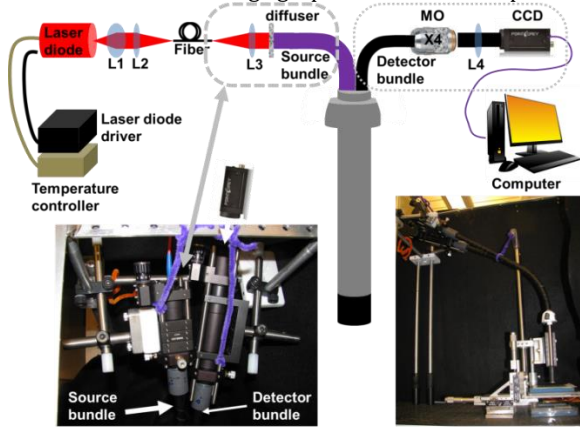

**Fig. 3.** A 785nm laser diode is coupled into a fiber through L1 and L2, for delivery to the probe-holding platform, where the light is collimated by a L3 and beam-homogenized by a diffuser for even illumination of the 70 source channels in a bundle. The 58 detector channels grouped in another bundle is projected to a CCD camera using a microscopic objective lens and a relay lens L4. The photographs show the rigid probe (lower right) and a close-up view of the optics/camera module (lower left) for light delivery to and light collection from the probe.

The method for reshaping the raw data to an *en-face* image is described in below. The 58 detector fibers within the bundle are radially interspersed (with the 70 source channels) at the tissue-imaging side but randomly positioned at the camera side. As a result, a raw image data acquired by the camera as shown in Fig. 4A contains the information of the light intensity collected by the 58 detector fibers at irregular orders. The circular region-of-interest (ROI) on the raw image data corresponding to each detector fiber-channel has an average radius of 60 pixels. The average intensity within a circular ROI of a radius of 20 pixels centered at each fiber position on the raw image data is calculated, as illustrated in Fig 4B. A blank matrix of 1448x1928 identical in matrix size to the raw image is produced with the vertical dimension representing 9.5mm—the diameter of the tissue-contacting fiber-probe. The average light intensity corresponding to each detector fiber position on the raw image as calculated from Fig. 4B is sort-mapped onto the entire area of a circular ROI in the blank matrix corresponding to the actual position of the detector fiber on the probe. The corresponding orders of the positions of the 58 detector fibers are shown in Fig. 4C. The pixel values for the ROI corresponding to each source fiber position of the 70 source fibers on the probe as numerated in Fig. 4(D) is assigned according to the method described in below.

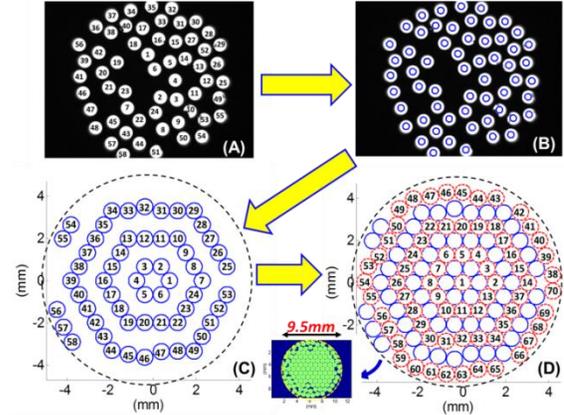

**Fig. 4.** (A) A raw data as an image acquired by CCD. (B) The average pixel intensity corresponding to each detection fiber is calculated. (C) The intensity of each detection fiber is mapped to the corresponding position of the fiber among the 58 detector fibers on the probe. (D) The pixel intensity corresponding to the position of each of the 70 source fibers are estimated by a weighting algorithm. The result is an image with a FOV of 9.5mm in diameter (see the smaller figure marked by the curved arrow left to D) that consists of 128 circles each representing a tissue area of 750um in diameter.

Denote $\psi_{homo}(\vec{r}_i^S, \vec{r}_j^D, \lambda)$ as the continuous-wave signal at a wavelength $\lambda$ from a homogeneous medium when illuminated by a source at $\vec{r}_i^S$ and acquired by a detector at $\vec{r}_j^D$. When all M sources illuminate simultaneously, the composite signal measured by one of the N detectors at $\vec{r}_j^D$ is $\psi_{homo}(\vec{r}_j^D, \lambda) = \sum_i \psi_{homo}(\vec{r}_i^S, \vec{r}_j^D, \lambda)$, $j = [1, N]$. Now assume that the source and detector channels could be swapped. Then when all of the N previously-detectors-now-sources are ON simultaneously, the hypothetical composite signal at one of the M previously-sources-now-detectors at the position $\vec{r}_i^S$ becomes $\psi_{homo}(\vec{r}_i^S, \lambda) = \sum_j \psi_{homo}(\vec{r}_i^S, \vec{r}_j^D, \lambda)$, $i = [1, M]$. The closer $\vec{r}_j^D$ is to $\vec{r}_i^S$, the more $\psi_{homo}(\vec{r}_i^S, \vec{r}_j^D, \lambda)$ contributes to $\psi_{homo}(\vec{r}_j^D, \lambda)$ and $\psi_{homo}(\vec{r}_i^S, \lambda)$; thus $\psi_{homo}(\vec{r}_i^S, \lambda)$ can be estimated as $\psi_{homo}(\vec{r}_i^S, \lambda) = \left(\frac{M}{N}\right) * \sum_j \psi_{homo}(\vec{r}_j^D, \lambda) W_{dist}^{(i,j)} / \sum_j W_{dist}^{(i,j)}$, where the weight W scales the contribution of $\psi_{homo}(\vec{r}_j^D, \lambda)$ to

$\psi_{\text{homo}}(\vec{r}_i^S, \lambda)$ according to the distance of $\vec{r}_j^D$ from $\vec{r}_i^S$. The combination of $\Psi_{\text{homo}}(\vec{r}_i^S, \lambda)$ and $\psi_{\text{homo}}(\vec{r}_i^S, \lambda)$ is denoted as $\psi_{\text{homo}}(\vec{r}, \lambda)$. When a near-surface anomaly $\Delta\mu(\vec{r}^{\blacksquare}, \lambda)$ is present, the composite signal changes to $\psi_{\text{hete}}(\vec{r}, \lambda)$. Then $\langle \psi_{\text{homo}}(\vec{r}, \lambda) - \psi_{\text{hete}}(\vec{r}, \lambda) \rangle = J(\vec{r}^{\bullet}, \vec{r}, \lambda) \cdot \Delta\mu(\vec{r}^{\bullet}, \lambda)$ according to the Born approximation, where $J$ is the sensitivity of the signal at $\vec{r}$ for an anomaly at $\vec{r}^{\blacksquare}$, and the magnitude of J peaks when $\vec{r}$ is the closest to $\vec{r}^{\blacksquare}$. Therefore a normalization of the signal versus a baseline as $1 - \Psi_{\text{hete}}(\vec{r}, \lambda)/\Psi_{\text{homo}}(\vec{r}, \lambda)$ (which also cancels the effect of the aforementioned weight W) qualitatively maps $\Delta\mu(\vec{r}^{\blacksquare}, \lambda)$. We thus assign a differential tissue heterogeneity index of $\Delta\mu(\vec{r}, \lambda) = \Psi_{\text{hete}}(\vec{r}, \lambda)/\Psi_{\text{homo}}(\vec{r}, \lambda)$ to the region under a fiber as the fiber-region pixel-value on the final *en-face* image.

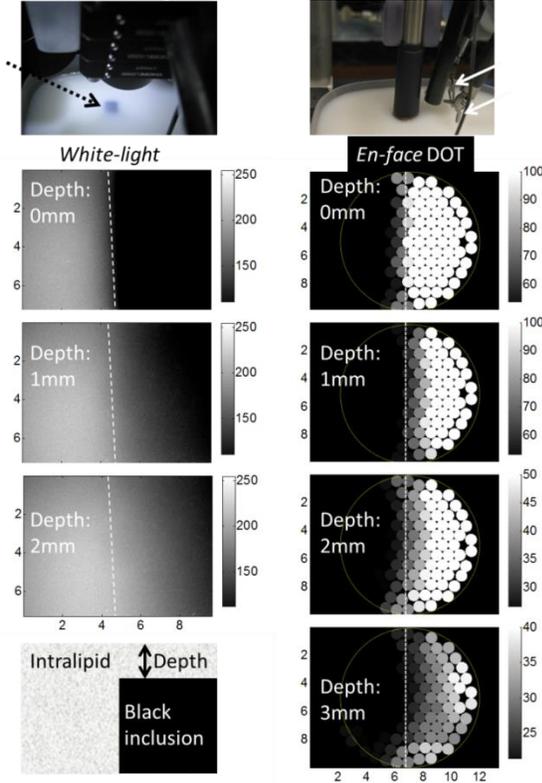

**Fig. 5.** (Left column) An 8mm square black object (pointed by the dashed black line) was embedded in 1% intralipid solution and took the right-half FOV of a white-light surface imaging system. It is difficult to assess object margin at a depth of 1mm. (Right column) The 8mm square black object was embedded in 1% intralipid solution and took the right-half FOV of the en-face laparoscopic applicator probe (robotic surgical instruments at the vicinity are marked by the two solid arrows). The presence of the margin formed by the black object can be appreciated at the object depth up to 3mm (see **Visualization 1**). Unit of the size dimension: mm. Contrast of the gray-scale: 2 folds.

An example of real-time *en-face* DOT of near-surface margin invisible to surface measurement of reflectance is given in Fig. 5. Surface imaging of white light reflectance was acquired over a rectangular FOV of 7.3mm×9.7mm, using another camera (PointGrey GigE vision, 1928×1448 pixels) and a 10X microscope objective lens. An 8mm square black object was embedded in 1% intralipid solution for lowering to different depths. The black object was positioned at the right-half of the camera FOV. On surface imaging of the white light reflectance the object margin was difficult to assess at a depth of 1mm. When imaged by *en-face* DOT, the margin formed by the black object could be continuously appreciated at a depth up to 3mm. A video footage of the object margin being revealed at a depth of 3mm as it was brought from a 5mm depth is available for review online.

Figure 6 is the *en-face* DOT visualization of a near-surface strip of light attenuation contrast formed by a needle-track of ink injection into a thermally coagulated avian egg tissue at an approximate depth of 1mm. En-face DOT resolved the orientation of the contrast-track, and indicated a sub-surface spatial extent of the attenuation contrast wider than was assessable at the surface that agreed with the injection mark.

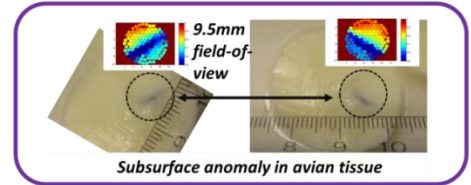

Fig. 6. En-face DOT of a near-surface strip of light attenuation contrast formed by needle-track of ink injection into a thermally coagulated avian egg white at an approximate depth of 1mm, at two orientations.

In summary, we demonstrated a method for real-time laparoscopic *en-face* topographic mapping of near-surface heterogeneity, by using an applicator probe housing 128 fibers that form radially alternating source and detector channels. By simultaneously illuminating the source channels and concurrently measuring the light diffusely propagated to the detector channels, real-time visualization of subsurface attenuation contrast at a depth up to 3mm invisible to surface reflectance imaging is demonstrated at a frame rate of 1.25Hz. This laparoscopically applicable method, when extended to probing spectral and fluorescence heterogeneities and using non-contact probe, may have the potential for intraoperative margin assessment during minimally invasive procedures.

**Funding.** Peggy and Charles Stephenson Cancer Center of the University of Oklahoma Health Sciences Center, Kerr Chair, Bullock Chair, and Oklahoma State University.